# Magnetothermal properties of molecule-based materials

M. Evangelisti,[a] F. Luis,[b] L. J. de Jongh[c] and M. Affronte[*a,d]

[a] National Research Center on "nanoStructures and bioSystems at Surfaces" ($S^3$), INFM-CNR, 41100 Modena, Italy
[b] Instituto de Ciencia de Materiales de Aragón, CSIC - Universidad de Zaragoza, 50009 Zaragoza, Spain
[c] Kamerlingh Onnes Laboratory, Leiden University, 2300 RA, Leiden, The Netherlands
[d] Dipartimento di Fisica, Università di Modena e Reggio Emilia, 41100 Modena, Italy

We critically review recent results obtained by studying the low-temperature specific heat of some of the most popular molecular magnets. Perspectives of this field are discussed as well.

## 1. Introduction

Supramolecular chemistry provides a potentially attractive way of assembling and maintaining controlled nanostructures. The magnitude of the magnetic interactions between the molecular building-blocks is crucial to determine the properties of such materials. The most intriguing situation is represented by molecular magnetic clusters in which a total spin can be defined for each individual cluster as a result of dominant intracluster interactions. The main parameter in this discussion is the *energy level separation* between different spin states that gets enlarged by the reduced size of the molecular units. Clearly at variance with bulk materials, for which energy levels constitute a continuum band, these systems can be considered in between the classical and quantum world.[1]

The specific heat $C$ of a given material, that is the heat capacity per unit mass (or mole), is defined as the derivative of its internal energy, $C = dU/dT$. Its determination thus provides direct information on the energy levels of the magnetic system. Because in a typical molecular cluster there are energy levels split in the range $1-10^2$ cm$^{-1}$ (1 cm$^{-1}$ = 1.43 K), anomalies in the specific heat are easily visible at liquid He temperature. One important feature that characterizes the specific heat technique with respect, e.g., magnetization or NMR, is that it can be employed in zero applied-field; no magnetic perturbation is therefore introduced in the system. On the other hand, an external magnetic field in the range $10^{-1}$–10 T gives rise to a Zeeman splitting of one or a few Kelvin to a molecular spin, so the anomaly in the specific heat can be consequently shifted in a controlled manner. In addition, the dependence of $C$ on magnetic fields gives information on the magnetic states, i.e., whether they are "classical" spin states or quantum superpositions (magnetic Schrödinger's cat states) of these. Specific heat experiments are currently carried out to detect *fine energy splittings* in molecular magnets in a versatile way and with an accuracy that is comparable to that of spectroscopic techniques.[2–7]

The approach to thermal equilibrium requires that energy is exchanged between the spins and the lattice, which constitutes the thermal reservoir. The so-called single-molecule magnets (hereafter SMM's)[1,8] relax to equilibrium by flipping the macroscopic spins over the classical anisotropy barrier if the thermal energy is sufficiently large. Below certain blocking temperature, the relaxation slows down drastically and quantum tunnelling through the anisotropy barrier is seen to contribute to the dynamics of the macroscopic spins. As we shall see below, specific heat experiments provide a privileged tool to investigate *time-dependent* phenomena in the quantum regime of such systems.[9–16]

Spin-spin correlations, that is *ordering phenomena*, are ideally investigated using specific heat experiments.[17] Given the fact that the interaction energies for molecular spins are usually weak, phase transitions may only occur at very low temperatures, typically below 1 K. In these cases, the application of any external -even small- magnetic field may wash out the collective phenomena and the use of a non invasive technique, such as the specific heat, is particularly suitable, if not necessary.[18,19]

The specific heat is also directly related to the magnetic entropy $S$ of the system through $S = \int (C/T) dT$. This quantity is currently receiving an increasing attention[20–23] for the search of possible applications of molecular magnets as *refrigerant materials*.

In what follows we address these issues by presenting selected examples, whose list is far from exhaustive, being not this the scope of the article. Rather, examples were chosen (i) to show some of the potentialities that this type of research has achieved, (ii) to draw the attention to open, but addressable, questions. The article is organized as follows. To provide a common language and references to the Reader, Sections 2 and 3 review the experimental techniques for specific heat measurements and the basic models, respectively. In Section 4, we discuss molecular antiferromagnetic rings, which yield to clear-cut manifestations of the Schottky anomaly easily tuned by external magnetic fields. Section 5 deals with the ability of the time-dependent specific heat technique to address the fundamental role played by phonons as well as nuclei in the spin-lattice relaxation deep in the quantum regime. Long-range magnetic correlations in high-spin molecular clusters are analyzed in Section 6 by means of specific heat experiments, with particular emphasis on dipolar magnets. Intimately related to the thermodynamics, the magnetocaloric effect of molecule-based materials is reviewed in Section 7, whereas the concluding Section 8 is devoted to probable future perspectives of this field.

## 2. Experimental determination of the specific heat

The essential calorimeter for low-temperature measurements of specific heat is made of a sample holder, typically one sapphire or Si slice, with attached a small thermometer and a heater that can be a simple metal evaporation or a small resistor (see Fig. 1). Four or more wires hold the calorimeter and they often constitute the electrical and the thermal link at the same time. For this reason, the size and the material of which these wires are made of need to be carefully chosen according to the temperature range and the size of the sample. In some cases, electrical connections to the calorimeter are ensured by superconducting wires, which inhibit heat transfer at the very low temperatures. Heat link from the thermal bath to the calorimeter is thus provided by an additional metallic wire. By changing this thermal link, the characteristic time of the experiment, that is the time needed to collect one single measurement, can be varied accordingly. Typical characteristic times vary between 1 and $10^2$ s, and are also dependent on temperature. The use of time-dependent specific heat experiments will be exemplified in Section 5.

The sample with heat capacity $C_s$ is thermally linked to the thermometer and heater through the sample holder. The empty calorimeter has a small but finite heat capacity referred as $C_{add} = C_{holder}+C_{thermo}+C_{heater}$, where *add* stands for *addenda*. In first approximation, we may suppose that these three objects are in excellent thermal contact and linked to the thermal bath through a controlled thermal impedance having thermal conductivity $K_1$. A controlled heat quantity $Q(t)$ is provided by the heater to the sample and the holder block so the response of the system is given in terms of temperature changes $T(t)$. The relevant quantity is the total heat capacity $C_{tot} = C_s+C_{add}$. The heat $Q(t)$ also flows to the thermal bath through the thermal conductivity of the wires so the energy balance for this system is:

$$\frac{\partial Q(t)}{\partial t} = K_1(T_1 - T_0) + C_{tot}\frac{\partial T_1}{\partial t} \quad (2.1)$$

where the term on the left is the rate of heat dissipated by the heater, the first term on the right is the heat flowing from the system to the thermal bath and the second term on the right is the heat used for the temperature variation of the system (sample + addenda). $T_0$ is the fixed temperature of the thermal bath while $T_1(t)$ is the actual temperature of the system (sample + addenda) that is time dependent. Different solutions of Eq. (2.1) are possible with a convenient choice of the thermal conductivity $K_1$ and of the time dependence of the heat $Q(t)$.

If $dQ(t)/dt = Q_0\delta(t-t_0)$ (heat pulse at $t = t_0$ for a very short time) and $K_1$ is very small (adiabatic condition), the solution is $T_1-T_0 = Q_0/C_{tot}$. Thus, controlling the value of $Q_0$ and measuring the temperature variation $T_1-T_0$, one may directly estimate the heat capacity $C_{tot} = Q_0/(T_1-T_0)$. This describes the *adiabatic methods* and it coincides with the definition of specific heat.

If $dQ(t)/dt = P\theta(t-t_0)$ (step like heat signal starting at $t = t_0$) and $K_1$ is finite, then the solution of Eq. (2.1) is $T_1-T_0 = PK_1^{-1}\exp(-(t-t_0)/\tau_1)$, where $\tau_1 = C_{tot}/K_1$ is the relaxation time. This method, known as *relaxation method*, was proposed by Bachmann et al.[24] and it is now widely used for automatized measurements.

If $dQ(t)/dt = p_0\cos^2(\omega t)$ (sinusoidal current through the heater) and $K_1$ is finite, then the steady state solution is $T_1-T_0 = p_0(4\omega C)^{-1}\alpha\sin(2\omega t)+p_0(2K_1)^{-1}$, where $\alpha$ is a correcting factors that in optimal condition is close to 1. This method was initially proposed by Sullivan and Seidel[25] and is known as the *ac method*.

Whenever the rate of internal equilibrium $\tau_2 = C/K_2$ becomes comparable with $\tau_1$ or $\omega$, the sample is not at thermal equilibrium with the thermometer during the heat change. Typical samples of molecular magnets have bad thermal conductivity $K_2$ and large specific heat. One thus has to consider a temperature gradient between the sample and the holder or within the specimen itself. In these conditions the analysis gets more complicated but approximated solutions of the energy balance are still possible under certain limits. We refer to the original work of Sullivan and Seidel[25] for a more sophisticate analysis of the *ac method* and we just remind that the correction factor $\alpha$ can be expressed as $(1-(8\omega^2\tau_1^2)^{-1}+2\omega^2\tau_2^2+K_1/K_2)$. In the case of the *relaxation method*, deep analysis and solutions were presented by J.P Shepherd[26] and by J.S. Hwang, K.J. Lin and C. Tien.[27] It turns out that, in the limit $\tau_1 > \tau_2$, the temperature rise (or decay) following a heat pulse is essentially given by the sum of two exponentials and the total specific heat can be analytically derived from the fit of the temperature relaxation with good precision, typically 1 or few per cent in the limit of $\tau_1 > \tau_2$. This is the reason why this method is currently used in automatic data acquisition systems. The control of the $\alpha$ coefficient for the *ac method* is less easy since it is very sensitive, for instance, to temperature changes. This method is indeed known to have accuracy not better than 10% or even worse under normal operative conditions. On the other hand the use of lock-in amplifiers allows precise detection of very small ac signal, thus the *ac method* is preferred when the knowledge of the absolute value is not essential and very small heat signals are requested (for instance to detect phase transition anomalies).

## 3. Basic models for the contributions to the specific heat

The specific heat of molecular magnets contains different contributions which can be understood within the framework of conventional solid state models as discussed in several textbooks at which the reader is referred for a deeper discussion.[28]

In a perfect crystal, there are $3N$ independent modes of lattice vibration, with $N$ the number of ions in a unit cell. Consequently, since molecule-based materials usually contain hundreds of ions per cell, their specific heats tend to saturate to a huge Dulong and Petit value $3NR$ at high temperatures. Even at cryogenic temperatures the lattice contribution is still substantial and, since it grows very fast with temperature, at $\approx 10$ K the lattice contribution $C_{latt}$ becomes already predominant, masking most of the features of the magnetic contribution $C_m$ thereby limiting the accuracy to which it can be determined. Furthermore, the details of the spectrum of the quantized lattice vibrations (phonons) can be quite complex due to the huge number of independent vibration modes and no detailed study of this type has been attempted so far for this class of materials. One usually applies the approximation of an idealized simplified crystal with three acoustic branches and ($3N$–3) optical branches. The three acoustic branches are treated within the Debye model, giving the so-called Debye contribution $C_D$, while the optical modes are described by the Einstein model giving the corresponding $C_E$ contribution, so that $C_{latt} = C_D + C_E$. Within the molecular clusters the atoms are strongly bonded, e.g. by ionic bonds, whereas intermolecular clusters are much weaker (often the clusters form molecular crystals through van der Waals bonding). Correspondingly, high- and low-energy branches are both present in the phonon spectra.[29] Optical modes correspond nearly to the vibration spectra of the isolated molecule, giving rise to Einstein modes that can be broadened in energy by intermolecular interactions. The Einstein and Debye contributions are respectively expressed as:

$$\frac{C_E}{R} = 3\left(\frac{T_E}{T}\right)^2 \frac{\exp\left(\frac{T_E}{T}\right)}{\left[\exp\left(\frac{T_E}{T}\right)-1\right]^2} \quad (3.1)$$

where $\varepsilon_E = k_B T_E$ is the energy of the optical phonon mode and

$$\frac{C_D}{R} = 9\left(\frac{T}{\Theta_D}\right)^3 \int_0^{\Theta_D/T} \frac{x e^x}{(e^x - 1)^2} dx \quad (3.2)$$

where $h\omega_D = hck_D = 2\pi k_B \Theta_D$ is the Debye energy for which the characteristic linear dispersion $\omega_D = ck_D$ of the acoustic branch is assumed, with $c$ the velocity of sound. For $T \ll \Theta_D$ (typically $T \leq \Theta_D/50$) the latter equation can be approximated as:

$$\frac{C_D}{R} = 234\left(\frac{T}{\Theta_D}\right)^3. \quad (3.3)$$

In practice the $T^3$ behaviour is hardly visible in molecular compounds unless one may neglect (or independently knows) the magnetic contribution at very low temperature. In the temperature range 0.1–20 K the typical behaviour of the lattice contribution is $C_{latt} \propto T^\alpha$ with $\alpha \sim 2.6$–3 and the typical $\Theta_D$ values that one obtains range between 15 and 50 K. These $\Theta_D$ values are much lower than usually found for insulating solids or intermetallic compounds ($\Theta_D$ typically ranges within 150 to 600 K in these cases) and also peculiar are the $T_E$ values that can be as low as 20–30 K. Both these features can simply arise from the weak intermolecular bonds. $T_E$ values can also arise from very localised optical modes and they are also found in disordered organic solid solvents. At any rate, the $\Theta_D$ values obtained cannot be expected to correspond to a simple unique sound propagation velocity associated with a single $\omega(k)$ dispersion relation, so one should use such low $\Theta_D$ values with care. Alternative phenomenological expressions to fit $C_{latt}$ are presented in Ref. [30] where the specific heat of a crystal composed of non-magnetic molecular rings was also studied.

The magnetic contribution $C_m$ to the specific heat is readily understood if one calculates the energy $E_m$ of a two- or multi-level magnetic system and assuming a canonical Boltzmann distribution describing the population of levels. So, since $C_m = dE_m/dT$, for a two-level system we have the expression of the well known Schottky anomaly:

$$\frac{C_S}{R} = \frac{g_0}{g_1}\left(\frac{T_0}{T}\right)^2 \frac{\exp\left(\frac{T_0}{T}\right)}{\left[1 + \frac{g_0}{g_1}\exp\left(\frac{T_0}{T}\right)\right]^2} \quad (3.4)$$

where $g_0$ and $g_1$ are the degeneracies of the ground and the first excited state, respectively, and $k_B T_0$ is the energy gap between these two states. If $g_0 = 1$ and $g_1 = 1$ the Schottky anomaly has a characteristic maximum at $T \sim T_0/2$ so the energy gap $k_B T_0$ can be immediately visualized by looking at the temperature at which the anomaly exhibits its maximum. If the magnetic system has many levels $E_i$ the Schottky expression can be generalized as:

$$\frac{C_S}{R} = \beta^2 \frac{\sum_i E_i^2 \exp(-\beta E_i)\sum_i \exp(-\beta E_i) - \left[\sum_i E_i \exp(-\beta E_i)\right]^2}{\left[\sum_i \exp(-\beta E_i)\right]^2} \quad (3.5)$$

This expression is commonly used to describe the magnetic contributions in molecular magnets. For instance if the distribution of the lowest lying energy levels is known, $C_S$ can be easily calculated or alternatively one may introduce one or few parameters describing the magnetic system and fit the specific heat data to fix them.

## 4. Specific heat of molecular antiferromagnetic wheels

Molecular rings have recently attracted considerable interest for several reasons: in many cases they have almost planar shape with an axial symmetry of the spin arrangement whilst the dominant antiferromagnetic Heisenberg exchange coupling between nearest neighbour spin centres and the even number of spins provide a singlet ground state at low temperature and zero field to the isometallic wheels. The lowest lying excitations follow a Lande's rule for which the energy gap between subsequent levels is proportional to $S(S+1)$ where $S$ is the spin of the excitation state whilst some other excitations forms further bands[31] with features related to the special boundary conditions of these finite AF spin chains. The appeal of these AF rings comes firstly from the energy separation between the lowest lying states that range between few Kelvin and few tens of Kelvin, so that the levels are well isolated at liquid helium temperature.

Moreover magnetic fields of few Tesla, easily accessible in the laboratory, may lower the energy of excited states leading to subsequent level crossing and spin flip of the ground state for increasing magnetic fields. Among the even membered cyclic spin systems, one of the first examples were the ferric wheels made of $Fe^{3+}$ ions carrying a $s = 5/2$ spin. Several isometallic molecular rings with different nuclearity exist, namely, $Fe_6$,[32] $Fe_8$, $Fe_{10}$,[33] $Fe_{12}$[34] and $Fe_{18}$.[35] Of great interest is also a family of cyclic systems made by $Cr^{3+}$ ions. The best example is a molecular ring containing eight $Cr^{3+}$ ions disposed in an almost perfect octagon (yet, only $C_4$ point group symmetry at room temperature).[36] This molecular compound has chemical formula $[Cr_8F_8(O_2CCMe_3)_{16}]$ and it crystallizes in a tetragonal or monoclinic space group with four halves of molecules per unit cell. Centimetre-sized single crystals can be grown and they are relatively stable in ambient conditions. *Ab-initio* calculations show that $Cr^{3+}$ ions carry a local spin moment $s = 3/2$ and that the preferred arrangement of the eight spins is antiferromagnetic.[37]

For what concerns the specific heat, the case of $Fe_{10}$ can be considered as paradigmatic[2] (Fig. 2). The scale of energies allows indeed in this case a clear identification of different contributions: at tens of Kelvin, the lattice dominates, at 1–3 K magnetic Schottky anomaly arises while the mK region is dominated by hyperfine contributions. It can be easily noticed in Fig. 2 that the Schottky anomaly, arising mainly from the energy gap between the ground state $S = 0$ and the first excited state with $S = 1$, becomes well resolved at low temperature while the lattice contribution dominates already above 6 K. The Schottky anomaly can be tentatively accounted for by considering simple two energy levels separated by an effective energy gap $\Delta/k_B = 4.47$ K: the exponential drop of $C/R$ at low temperature is well described by such a simple model, yielding an estimate of the energy separation between the ground state and the barycentre of the first excited triplet. At higher temperatures, further excited states cannot be neglected and the magnetic contribution must be separated from the lattice contribution for a correct analysis. At 0.5 K a small bump, probably due to low energy excitations in a small fraction of defected rings, is visible and below 0.3 K, $C/R$ shows an upturn that scales as $T^{-2}$. The latter is due to hyperfine interactions. We may fit the specific heat data to determine the characteristic parameters of the spin Hamiltonian. The position of the Schottky anomaly provides a precise determination of the lowest energy gap without the need for applying a (perturbing) external magnetic field and no need of using single crystals. This is, in general, enough to fix the main Heisenberg exchange coupling in the limit in which this is the leading term of the spin Hamiltonian, but only a rough estimation of the zero field splitting parameter can be given since this does not give any distinctive feature (only a broad contribution) to the $C(T)$ curve.

A way to overcome this problem is to apply external magnetic fields and measure a few $C(T)$ curves for different $H$ values. In this case the energy level scheme changes as a consequence of the Zeeman term and the Schottky anomaly shifts consequently. In this way different independent $C(T,H)$ curves can be used to determine the microscopic parameters of the spin Hamiltonian. The latter can be exactly diagonalized if the ring has few ($n$) centres and/or their spin ($s$) is low. Yet, for $n > 10$ and $s \geq 5/2$ the Hilbert space is already huge $[(s)^n]$ and the exact diagonalization becomes unfeasible even with the modern computational facilities. Nevertheless, the way in which the physical properties of finite spin chain merge to those of an infinite chain is an interesting issue and theoretical methods to treat such mesoscopic systems deserve much attention since they are of fundamental importance to describe nano-objects. A semiclassical model was proposed to describe AF rings made of generic $n$ spin centres and this provides an analytical solution for the lowest lying states[38] that was used to fit the specific heat of $Fe_{12}$ ferric wheel.[3]

This semiclassical model was also used to describe the tunnelling of the Néel vector, which phenomenon is analogous to the tunnelling of the cluster spins observed in high-spin $Mn_{12}$ and $Fe_8$, except that the net total spin of the (ferrimagnetic) molecule is replaced by the Néel vector of the antiferromagnetic molecule. This phenomenon was recently discussed and searched for in molecular rings. The first problem in obtaining clear evidence for such a tunnelling is related to the fact that the Néel vector gives no magnetic signal to be detected with conventional magnetometers. The second issue is the identification of a specific energy barrier through which the Néel vector can tunnel. This is a subtle point since by sweeping the magnetic field the AF spin system switches through different spin wave modes that do not necessary imply a true tunnelling of the Néel vector through an energy barrier. Such a barrier and the corresponding tunnelling events must be carefully determined from the spin Hamiltonian. Exact diagonalization methods are preferable for this task although the semiclassical model provides the essential physics. Specific heat measurements can be performed at fixed temperature ($T \leq 1$ K) and in a sweeping magnetic field in order to determine precisely the energy barrier involved in the tunnelling of the Néel vector. The debate is still open on this interesting topic.

Specific heat measurements at fixed temperature and high magnetic field also allow us to address further interesting issues. As previously mentioned, in the case of homometallic rings the first excited state $S = 1$ is lead to cross the ground state, namely the $S = 0$ singlet, by an external magnetic field of a few Tesla. Increasing the field even further, the new ground state (with quantum number equal to $S = 1$) is progressively crossed by subsequent excited (i.e. $S = 2,3,...$) states. For symmetric rings the two lowest lying excited states, with $S = 1$ and $S = 2$ quantum numbers, have different parity, i.e. their wave functions have different irreducible representations. It turns out that terms of the spin Hamiltonian with the axial symmetry of the ring are not able to mix the $S$ and the $S+1$ states, so a true crossing is expected in this case with no repulsion between states. An odd case was presented in Ref. [4] in which the specific heat of $Fe_6$ rings was measured at 0.78 K in sweeping field up to 28 T. These measurements did not necessarily require a high precision of the absolute value of $C$, but a high sensitivity to measure the changes of $C$ as the external magnetic field is swept. For such reasons, the *ac* measuring method was preferred in this case. A first crossing between the $|0\rangle$ and the $|1,-1\rangle$ was observed at the crossing field $B_{c1} = 11.7$ T and a second crossing between the $|1,-1\rangle$ and the $|2,-2\rangle$ at $B_{c2} = 22.4$ T. Close to the crossing field the system is well represented by the two lowest levels only. The Schottky anomaly can be expressed as $C = (\Delta/k_BT)^2 \exp(\Delta/k_BT)[1+\exp(\Delta/k_BT)]^{-2}$ and it exhibits maxima when the energy gap $\Delta \sim 2.4\ k_BT$ while it is expected to vanish for $\Delta = 0$. In these experiments the specific heat did not vanish at $B_c$'s indicating a level repulsion, contrary to what is expected for true crossing between two levels. The origin of this behaviour can be due to the presence of anti-symmetric terms in the spin Hamiltonian, like the Dzyaloshinski-Morya one, which do not have the axial (i.e. $S_6$ point group) symmetry of the Fe$_6$ ring. These terms can mix states with different parity and then induce the level anticrossing.[5] Yet, other extrinsic factors, like inhomogeneity within the crystals, may lead to broadening of the heat capacity peak and eventually to non vanishing heat capacity at $B_c$'s. These experiments were repeated on molecular Cr$_8$ rings[39] and in this case the Schottky anomaly was found to vanish at the crossing fields within the experimental accuracy[6] (Fig. 3). These experiments showed, at least, the great potentiality of specific heat measurements to get a direct estimate of the repulsion between two crossing levels on bulk samples.

A step further was done as soon as heterometallic molecular rings were synthesized. In particular one of the eight Cr ions was replaced by a divalent transition metal, like Zn, Ni, etc. Thus an "extra" spin was introduced in the otherwise compensated Cr$_8$ ring and the ground state in zero field is no more a singlet. The axial symmetry of the ring was broken and consequently two consecutive states, with quantum number $S$ and $S+1$, can easily be mixed by terms in the spin Hamiltonian such as the anisotropic one. In this case the mixing of two consecutive states is enhanced at the crossing field and the total spin of the ring oscillates between two states with quantum number $S$ and $S+1$. These quantum oscillations can be tuned by changing the angle $\theta$ between the ring axis and the external magnetic field and the level repulsion vanishes for angle $\theta$ close to zero and $\pi/2$ while an energy gap develops for intermediate orientations. This expectation was confirmed by specific heat experiments on heterometallic Cr$_7$Zn rings which have a ground state $S = 3/2$ in zero field, as shown in Fig. 4.[7]

## 5. Quantum tunnelling and quantum coherence in single-molecule magnets

The phenomenon of tunnelling is one of the most fascinating predictions of Quantum Mechanics, and one that fully contradicts our daily perception of the way in which "real" objects behave. Quantum tunnelling is directly associated with the description of a physical system in terms of wave functions and distributions of probability, which emanates from Heisenberg's uncertainty principle. By tunnelling across an energy barrier, a physical system, like a particle, can escape from a potential energy well towards a more stable state or even oscillate between equivalent configurations of equal energy. The first process explains, for example, the alpha decay of heavy nuclei,[40] which takes place when a nucleus of $^4$He irreversibly (or *incoherently*) escapes from the interior of an unstable nucleus. The second situation describes the *coherent* oscillation of the hydrogen plane of the ammonia molecule NH$_3$.[41] In the latter case, energy eigenstates are no longer associated with definite atomic locations (i.e. to the hydrogen plane being either below or above the N) but to linear superpositions of these. In addition, the classical degeneracy is lifted by an energy splitting $\Delta$, usually called the "tunnel splitting", between the energies corresponding to either symmetric or antisymmetric superpositions. The existence of a tunnel splitting $\Delta$, thus also of these weird quantum states, can be shown by specific heat measurements because it leads to a Schottky anomaly, Eq. (3.4), with a maximum for $\Delta = 2.4\ k_BT$. The ability of specific heat experiments to measure small energy splittings has been used to study microscopic systems, like molecules and defects embedded in solids, which perform hindered rotational or translational movements also by tunnelling.[42]

The single-molecule magnets are ideal materials to investigate these tunnelling phenomena at the mesoscopic scale, intermediate between the microscopic and macroscopic worlds. These clusters contain a core, made of a few magnetic ions strongly coupled to each other by superexchange interactions, surrounded by a shell of organic ligands. Crystal-field effects usually lead to a strong anisotropy for the net molecular spin $S$, which in the simplest uniaxial case favours two opposite orientations along a given molecular axis $z$. In a magnetic field and at sufficiently low temperatures, SMM's can be described by the following Hamiltonian[43]

$$\mathcal{H} = -DS_z^2 + \mathcal{H}' - g\mu_B(H_xS_x + H_yS_y + H_zS_z) \qquad (5.1)$$

The angular dependence of the magnetic potential energy has therefore the shape of the well-known double-well potential (see Fig. 5). Tunnelling across the potential energy barrier is allowed by terms that do not commute with $S_z$, such as those contained in $\mathcal{H}'$, associated with deviations from perfect uniaxial symmetry, or the off-diagonal Zeeman terms.

For objects as large as these molecules, though, the observation of tunnelling faces two important difficulties. The first arises from the fact that the tunnelling probabilities and $\Delta$ depend exponentially on the height $U_{cl} \approx DS^2$ of the energy barrier, which usually scales with the volume. The ground state $\Delta$ ($\approx 10^{-6}$ K for Mn$_4$, $10^{-8}$ K for Fe$_8$, $10^{-10}$ K for Mn$_{12}$) is completely unmeasurable by means of specific heat experiments. But, in addition, even at zero-applied magnetic field, the weak dipolar interaction with neighbouring molecular spins in the same crystal induces a bias $\xi = 2g\mu_BSH_{dip,z}$ that is many orders of magnitude larger than $\Delta$. The splitting of the ground state doublet then approximately equals $\xi$ and the lowest energy eigenstates are the classical spin-up or spin-down states.[44,45] Under these conditions, *incoherent* tunnelling can still provide the spins a mechanism for reversing their orientations and, as shall be considered in detail below, for the attainment of thermal equilibrium.

An important advantage of some SMM's is their tendency to form molecular crystals, in which the anisotropy axes of all molecular clusters are aligned. The off-diagonal energy terms in Eq. (5.1) can then be modified by the application of external magnetic fields $B_\perp$ *perpendicular* to the anisotropy axis $z$.[46] As Fig. 5 shows schematically, a magnetic field applied along the $x$ direction lowers $U_{cl}$ but does not affect the inversion $z \leftrightarrow -z$ symmetry of the double well potential. Since off-diagonal terms play the role of a kinetic energy in this tunnelling problem, this enables reducing the effective mass. By increasing $B_\perp$, it is then possible to significantly increase $\Delta$ (see the upper panel of Fig. 6), which can eventually become larger than the splittings $\xi$ caused by external perturbations. Under such conditions, the energy splitting of the magnetic ground state becomes dominated by $\Delta$ and the eigenstates become quantum superpositions of spin-up and spin-down states.

The lower panel of Fig. 6 shows specific heat data measured on an oriented sample of $Fe_8$ molecular clusters. Near 1.7 T, the data show a jump that indicates that only above this field the spins are in thermal equilibrium with the lattice.[47] The equilibrium specific heat shows a maximum centred on 2.5–3 T. The position of the peak agrees with the perpendicular fields for which $\Delta$ is expected to become of order of the thermal energy, as expected for the Schottky anomaly of a two-level system, Eq. (3.4). These field-dependent specific heat experiments neatly show the existence of a finite tunnel splitting $\Delta$. The results agree well with the specific heat calculated using the quantum energy levels of $Fe_8$. This is in sharp contrast with the behaviour expected for classical magnetic moments, whose spin-up and spin-down states remain degenerate in a transverse magnetic field.

These and similar results obtained for other SMM's, such as $Mn_4$ and $Mn_{12}$,[11,12,14] show that the application of transverse magnetic fields provides a very useful method to make quantum states robust against decoherence and how specific heat experiments enable us to measure the tunnel splittings. However, much work is required yet to control the *coherent* evolution with time of these superpositions, which could open the door for applications in the domain of quantum computation.

**Towards the thermal equilibrium of the molecular spins with the lattice**

The specific heat directly reflects the population of the different energy levels and how they are modified by temperature. Under thermal equilibrium conditions, populations are simply given by Boltzmann statistics. However, the presence of energy barriers separating magnetic molecular states can lead to a slow relaxation towards this equilibrium state, especially at low temperatures. As a result, the populations and therefore the specific heat depend on the timescale of the measurement.[9] Usually, the time-dependent magnetic specific heat can be approximated by an exponential decay[10]

$$C_m(t) = C_m^0 + (C_m^{eq} - C_m^0)\, e^{-\Gamma t} \qquad (5.2)$$

where $C_m^0$ is the contribution to the specific heat of "fast" re-equilibrating levels, $C_m^{eq}$ is the equilibrium specific heat, and $\Gamma$ is the relaxation rate. For SMM's having a double-well potential energy landscape, $C_m^0$ is usually associated with transitions between levels located on each side of the energy barrier, whose relative populations equilibrate at very fast rates $\Gamma_0 \sim 10^{-8}$ s. By contrast, transferring population between levels located on different potential energy wells requires crossing the energy barrier $U_{cl}$ and therefore takes place at a much lower rate $\Gamma$.

Measuring specific heat at varying experimental times can therefore be used to obtain information on the process, either classical or by quantum tunnelling, by which the molecular spins attain thermal equilibrium with the phonon lattice. It is important to mention also that, whereas any tunnelling process gives rise to a magnetization change, only those that imply a transfer of energy between the spins and the lattice contribute to the spin-lattice relaxation rate $\Gamma$. We next consider the different mechanisms that dominate the spin-lattice relaxation of SMM's as temperature decreases.

**Phonon-induced quantum tunnelling via thermally activated states**

Figure 7 shows $C_m$ of $Fe_8$ clusters measured at zero field as a function of temperature. The zero-field splitting caused by the anisotropy terms in Eq. (5.1) is of order 28 K for this cluster.[44,48] It gives rise to a multi-level Schottky anomaly, Eq. (3.5), with a maximum above 3 K. This anomaly corresponds to the intrawell contribution $C_m^0$. When $k_B T \ll (2S-1)D \approx 5.5$ K, it decays exponentially indicating the depopulation of all levels above the ground state doublet.

For lower temperatures, the specific heat is dominated by transitions between the two lowest lying magnetic states $\pm S$, split by dipolar interactions. However, as the Figure 7 shows, $C_m$ deviates from equilibrium below approximately 1.3 K.[11] By increasing the experimental time, the cross-over between equilibrium and off-equilibrium conditions shifts towards lower temperatures. The same behaviour is observed for other clusters, such as $Mn_{12}$.[11] It clearly indicates that $\Gamma$ depends strongly on temperature. Studies of the ac susceptibility and magnetic relaxation, have indeed shown that $\Gamma$ follows the Arrhenius law:[49-51]

$$\Gamma = \Gamma_0 \exp(-U/k_B T) \qquad (4.3)$$

This thermally activated process is a generalization of the Orbach process of paramagnetic atoms[52] and it is reasonably well understood.[53–56] Quantum tunnelling takes place predominantly via the lowest lying states $\pm m$ for which tunnelling is not blocked by dipolar interactions ($\Delta_m > \xi_m$). Since an external magnetic field applied along $z$ also induces a bias, this process is *resonant*, i.e., it becomes faster at the crossing fields $B_n \approx nD/g\mu_B$, when states located on opposite sides of the energy barrier become degenerate, allowing tunnelling to occur. This quantum effect leads to maxima in the non-equilibrium $C_m$ vs. $B_{||}$ curves, as is shown in Fig. 8.

If, by contrast, the field is applied perpendicular to the anisotropy axis, it increases $\Delta_m$ of all levels. Tunnelling takes place via progressively lower lying states, which gives rise to a gradual decrease of $U_{cl}$ and of the blocking temperature (see Fig. 7). Eventually (for $B_\perp > 1.7$ T in the case of Fe$_8$), relaxation takes place via direct processes,[52] i.e. by phonon-induced tunnelling between the two lowest lying states. The fact that this phenomenon occurs at rates of order 1 s$^{-1}$ indicates that the wave-functions of these two states overlap significantly. Under these conditions, $\Gamma$ depends weakly on $T$ and the spins remain in equilibrium down to very low temperatures, as we had seen in the previous section (Fig. 6) and it is shown again in Fig. 7.

**Spin-lattice relaxation in the quantum regime**

The mechanisms responsible for the spin-lattice relaxation at sufficiently high temperatures or applied magnetic fields are relatively well understood. They are, in fact, generalisations of the direct and Orbach mechanisms that operate in the case of paramagnetic atoms. Let us consider next the situation in the limit of very low temperatures (typically $T < 1$ K) and $B = 0$, when only the two lowest lying levels are populated. Since $\xi \gg \Delta$ for these states, coherent spin tunnelling is nearly forbidden and the direct phonon-induced processes have extremely small probabilities (of order $10^{-9}$ s$^{-1}$ for Mn$_4$ clusters and $10^{-14}$ s$^{-1}$ for Fe$_8$). It is however by now firmly established that tunnelling is made possible by the interaction with rapidly fluctuating hyperfine fields, which bring a significant number of electron spins into resonance.[57] Coupling to a nuclear spin bath allows ground state tunnelling over a range of local bias fields $\xi$ much larger than the tunnel splitting $\Delta$. Within this theory, *magnetic* relaxation could in principle occur with no exchange of energy with the phonons of the molecular crystal.[58] The question of whether this tunnelling process is able to bring the spins to thermal equilibrium is not yet fully understood.[13,15,16,58] As we show next, specific heat experiments are ideally suited to investigate this fundamental question from an experimental point of view.

Figure 9 shows the time-dependent specific heat of Mn$_4$Cl clusters.[15] Again, the specific heat is dominated, above 2 K, by the Schottky associated with the anisotropy and by phonon contributions. Below 1 K, $C_m$ deviates from equilibrium. These small clusters have relatively low anisotropy barriers $U_{cl} \approx 14$ K. This enables to measure a large fraction of the evolution of $C_m$ with time in the available experimental time window. By fitting Eq. (5.2) to this data, it is possible to extract $\Gamma$. This is shown in the lower panel of Fig. 9 as a function of temperature, where data obtained from magnetic relaxation experiments are also included. A cross-over from a thermally activated regime, dominant at high temperatures, to a temperature-independent one is evident from these data. This shows that spins are indeed able to attain thermal equilibrium via incoherent tunnelling processes. Remarkably enough, the measured $\Gamma$ is six orders (!) of magnitude larger than the rate predicted for conventional phonon-induced direct processes.[10,54] Also surprising is the fact that the rate extracted from specific heat experiments agrees quantitatively with what is obtained from the decay of the magnetization data.[58] In other words, the same tunnelling mechanism, mediated by nuclear spins, that flips the spins brings them also to thermal equilibrium. Contrary to the predictions of the model by Prokofe'v and Stamp,[57] the relaxation of the energy of the electronic spins takes place towards the phonon bath and not just to the nuclear spin bath.

The interpretation that the spin-lattice relaxation is mediated, in the quantum regime, by the interaction with nuclear spins can be tested by comparing results obtained for samples with different isotopic content. In Fig. 10, we show data measured for two samples of Fe$_8$ clusters.[16] The first was prepared with Fe containing mainly the nonmagnetic $^{56}$Fe nuclei, whereas the second one was enriched in the magnetic $^{57}$Fe isotope,[60] which carries a $I = 1/2$ nuclear spin. Above 1 K, the specific heat of both compounds is nearly the same, showing that the magnetic anisotropy and intermolecular vibrations are very similar for both. By contrast, the isotopic effect is evident at very low temperatures: for approximately the same time scales, the inclusion of nuclear spins leads to a much larger specific heat for the $^{57}$Fe$_8$.

As for Mn$_4$, the relaxation rate of $^{57}$Fe8 was extracted by fitting the variation of $C_m$ with experimental times.[16] The lower panel of Fig. 10 shows the rates deduced from different experimental techniques: ac-susceptibility, dc-magnetization,[60] NMR,[61] Mössbauer,[62] besides the specific heat.[11,16] Together, they result into a $\Gamma(T)$ curve whose frequency window spans remarkably over twelve orders of magnitude. It is seen that $\Gamma$ follows the thermally activated behaviour down to approximately 1 K, whereas below this temperature it becomes much less dependent on $T$. The rate extrapolated to zero-temperature is nearly ten (!) orders of magnitude faster than what would be expected for conventional phonon-assisted processes. Unfortunately, $C_m$ of the standard Fe$_8$ sample was too small to allow any determination of $\Gamma$. Yet, the measured data are compatible with $\Gamma$ being at least three times smaller than in the case of $^{57}$Fe$_8$. These results confirm the isotopic effects on quantum tunnelling that has been studied by magnetic relaxation experiments[60] and NMR.[61] In addition, it confirms that nuclear spins play a crucial role in the quantum mechanism of spin-lattice relaxation of the molecular spins, something that had not been predicted by any theory. It is curious that nuclear spins are themselves not in equilibrium,[62,63] as it is clear from the data of Figs. 9 and 10, since they exchange energy with the phonon bath via the interaction with the electronic spins. These results call for an extension of the nuclear-mediated tunnelling mechanism[57] to incorporate the possibility of emission and absorption of phonons.

## 6. Long-range magnetic ordering

Critical phenomena are often studied by means of specific heat experiments.[17] A major reason is the appearance of a λ-type anomaly in the specific heat at the critical temperature, in which the material undergoes an order-disorder phase transition. Magnetic susceptibility shows a similar behaviour but whereas for specific heat the λ-type anomaly implies a discontinuity in the derivative of the internal energy and therefore a phase transition (although it may be not magnetic), for the magnetic susceptibility it is not strictly so, and spikes may have other sources, i.e. blocking or freezing of the magnetic moments.

Long-range magnetic order (LRMO) in molecule-based compounds was searched since the late 80ies by exploiting the exchange pathways established between organic radicals. This research activity led, for instance, to the discovery of LRMO at room temperature in Prussian blue analogues.[64] In this class of materials, molecular building blocks couple to form an extended 3d network. Conversely, high-spin molecular clusters are usually isolated from each other by a shell of organic ligands that surround their magnetic cores. In such a scenario, cluster spins in molecular crystals replace what ions are in conventional materials. Little is known, however, about the collective magnetic behaviour of these nanostructured materials. Do they show phase transitions as conventional magnets do? Magnetic correlations between the clusters are established by long-range dipole-dipole interactions. The dipolar energy per molecular spin of an ordered crystal is given by

$$E_{dip} = \frac{R}{2mk_B} \sum_{i=1}^{m} \vec{S}_i \cdot \vec{H}_i, \quad (6.1)$$

where $m$ is the number of molecular spins per magnetic cell, and $H_i$ is the molecular field at site $i$, given by

$$\vec{H}_i = -\sum_{j \neq i}^{N} \left\{ \frac{\vec{S}_j}{r_{ij}^3} - 3\frac{(\vec{S}_i \cdot \vec{r}_{ij})\vec{r}_{ij}}{r_{ij}^5} \right\}. \quad (6.2)$$

In the limit of zero magnetic field and very-low temperature (typically $T < 1$ K), dipole-dipole interactions between the molecular spins may induce a phase transition to LRMO.[13,65–67] Much of the interest of current research in magnetic ordering is focused on this possibility because only a few examples of purely dipolar magnets are known to date.[18,19,68,69] It is worth noting that different critical behaviours are expected depending on the nature of the magnetic correlations. Assuming point-like magnetic units, dipole-dipole interactions between them are calculated without involving any adjustable parameter. Therefore, the macroscopic properties of dipolar magnets can be precisely predicted.[13,65–67] Such ideal materials are however very difficult to obtain. As often is the case, intermolecular superexchange interactions may not be negligible at the very low temperatures where the LRMO takes place. The consequence is that correlations between the molecules are often established by quantum-mechanical superexchange interactions at short ranges, whose macroscopic prediction is made difficult by their strong dependence on electronic details. Indeed, intermolecular superexchange interactions were found to be responsible for the observed LRMO in the high-spin molecular clusters Fe$_{19}$,[70] Mn$_4$Br,[71] Mn$_4$Me,[15] Fe$_{14}$,[22] while they probably also play a major role in the case of Mn$_{12}$.[72]

The absence of any superexchange pathway between the molecules is not the only pre-requisite needed for the experimental observation of purely dipolar LRMO. An obvious requirement is that molecules should have large molecular spins to lead to accessible ordering temperatures. Another complication is added by the cluster magnetic anisotropy. As we have shown in Section 5, the spin-lattice relaxation of clusters with large magnetic anisotropy, such as Mn$_{12}$ or Fe$_8$ becomes very slow below the blocking temperature $T_B$. Furthermore, dipole-dipole interactions make the approach to equilibrium a collective, thus even slower, process.[13,58] This situation makes it very difficult to investigate the equilibrium response. In addition, the cluster anisotropy energies, favouring the molecular spin alignment along dictated directions, compete with the intermolecular coupling. Although it has been shown theoretically[13] that the occurrence of quantum tunnelling between opposite spin directions at temperatures below $T_B$ may in principle produce sufficient fluctuations to overcome blocking, in most of the investigated anisotropic molecules the times involved for the actual observation of the ensuing LRMO are still much too long. The only exception reported so far is represented by the Mn$_4$Me molecular magnet which undergoes a transition to LRMO below $T_C = 0.21$ K,[15] as depicted in Fig. 11. In comparison with Mn$_4$Cl reviewed in Section 5, Mn$_4$Me has a lower symmetry,[15] which results in a larger tunnel splitting $\Delta_m$, and therefore in an enhanced spin-lattice relaxation rate. Specific heat experiments revealed indeed that for Mn$_4$Me the thermal equilibrium is maintained down to the lowest temperatures, enabling the observation of LRMO. Numerical simulations showed that the Mn$_4$Me molecules are correlated by weak superexchange interactions besides the dipolar ones.[15a]

The obvious and "simplest" way to look for LRMO purely driven by dipolar interactions, is to search for high-spin molecular clusters with as low anisotropy as possible. The Mn$_6$ molecular cluster belongs to this class of materials.[18] The magnetic core of the cluster is a highly symmetric octahedron of Mn$^{3+}$ ions ferromagnetically coupled by strong superexchange interactions. The result is that Mn$_6$ has a $S = 12$ spin ground state at low-temperatures and very small magnetic anisotropy $D \leq 0.05$ K.[18] Correspondingly, the classical potential energy barrier of Mn$_6$ amounts to $U_{cl} \approx DS^2 = 1.5$ K only. This value has to be compared with $U_{cl} \approx 70$ K and 30 K found for Mn$_{12}$ and Fe$_8$. We expected, and observed indeed, that the spin-lattice relaxation of Mn$_6$ remained sufficiently fast to enable specific heat measurements down to the ordering temperature. A sharp peak in the specific heat was reported[18] at $T_C = 0.15$ K. The experiment agreed well with the specific heat calculated for a $S = 12$ Ising model of magnetic dipoles performed on a lattice magnetically equivalent to Mn$_6$. The model included dipolar interactions as well as the cluster anisotropy. This simulation showed that the ground state of Mn$_6$ is ferromagnetically ordered.

A step forward in the search of dipolar magnets is represented by the Fe$_{17}$ molecular cluster,[19,73] which contains 17 Fe$^{3+}$ ions per molecule linked via oxygen atoms. Similarly to Mn$_6$, Fe$_{17}$ carries very-large spin ground state and small cluster anisotropy. These values are $S = 35/2$ and $D \cong 0.02$ K. respectively. In addition, molecules are bound together in the crystal only by van der Waals forces prohibiting thus any intermolecular superexchange pathway. What makes Fe$_{17}$ an unique model system is that, by changing the crystallization conditions, it is possible to change the molecular packing without affecting the individual molecules, that is keeping the surrounding ligands, the molecular high-spin ground state and magnetic anisotropy unaltered. In other words, with $Fe_{17}$ it becomes feasible to tune the dipolar coupling between molecules with respect to the single-molecule properties. The resulting interplay gives rise to macroscopic behaviours ranging from superparamagnetic blocking to LRMO.[19]

As an example, Figure 12 shows the specific heat $C$ collected at several applied-fields for two $Fe_{17}$ compounds, $Fe_{17}$-trigonal and $Fe_{17}$-cubic, having trigonal and cubic symmetries, respectively. It can be noticed that both compounds depend on the applied-field in an identical manner regardless of the trigonal or cubic symmetry. This implies that the $Fe_{17}$ magnetic molecule (that is the spin ground state and anisotropy) is the same in both systems, as corroborated also by magnetization data.[19] The main difference in Fig. 12 is in the zero-applied field data for which a λ-type anomaly centred at $T_C = 0.81$ K is observed for trigonal symmetry (inset of Fig. 12). This feature reveals the onset of LRMO, the magnetic nature is indeed proven by its disappearance upon application of $H$. Clearly, the λ-type anomaly arises on top of a much broader one, which shifts with increasing applied-field towards higher temperatures. Because of the small anisotropy ($D \cong 0.02$ K), it is expected that the magnetic contribution to $C(T,H)$ for $\mu_0 H \geq 1$ T is mainly due to Schottky-like Zeeman splittings of the otherwise nearly degenerate energy spin states. Indeed, the calculated Schottky curves (solid lines) arising from the field-split levels accounts very well for the experimental data. The same behaviour is followed by $Fe_{17}$-cubic except that no sign of LRMO is apparently observed. As particularly evident in the low-$T$ / high-$H$ region of Fig. 12, phonon modes contribute differently to $C(T)$ of $Fe_{17}$-trigonal and $Fe_{17}$-cubic (dotted lines), reflecting the different geometries of the crystal lattice. For $Fe_{17}$-trigonal, numerical estimates of the dipolar energy $E_{dip}$ agree well with the observed ordering temperature $T_C = 0.81$ K proving that the ordering is here driven by dipole-dipole interactions. Contrary, the symmetry of the molecules in $Fe_{17}$-cubic is such that the calculated $E_{dip}$ does not exceed the expected blocking temperature of a superparamagnet with $S = 35/2$ and $D \cong 0.02$ K. We indeed note that the energy barrier of the $Fe_{17}$ molecule is about eight times smaller than that of $Mn_{12}$.[8] One would expect, therefore, a blocking temperature of $T_B$ ($Mn_{12}$)/8 $\cong 0.5$ K, which indeed is what is observed for $Fe_{17}$-cubic by means of magnetic relaxation measurements.[19]

## 7. The magnetocaloric effect

Molecule-based materials compete at low-$T$ with lanthanide and intermetallic compounds conventionally employed as magnetic refrigerant materials. This is the scenario emerging from recent experimental results,[20,21,22] which show that the magnetocaloric effect (hereafter MCE), that is the cooling or heating in varying magnetic fields, may be enhanced in molecule-based materials. The MCE and the associated principle of adiabatic demagnetization are readily understood from the top panel of Figure 13. Besides the fundamental interest on the MCE properties of novel materials, this effect is of technological importance since it can be used for cooling applications[74] according to a process known as adiabatic demagnetization.[75] Before presenting any result, we point out that specific heat experiments are particularly suitable to study the MCE, since this effect is intimately related to the thermodynamics.[76]

Nanostructured materials made of superparamagnetic clusters are particularly appealing in terms of MCE.[77] This is because the usually large magnetic moments of the clusters are easily polarized by the applied-field providing large magnetic entropy changes. The picture depicted in Fig. 13 is still valid for a superparamagnet with $S$ as net cluster magnetic moment, provided that it is at temperatures above the blocking temperature. The cluster magnetic anisotropy, which indeed determines the blocking temperature, can be considered as a drawback in the MCE efficiency of superparamagnets. Let us suppose to have three systems of non-interacting mono-disperse magnetic clusters with spin $S = 10$ and hypothetical axial anisotropies $D = 0.5, 1.5$ and $3.0$ K, respectively. Thus, we first calculate the corresponding Schottky anomalies and then the magnetic entropy changes for a given field change. The results for $\Delta S_m$ are depicted in the lower panel of Fig. 13, where it can be noticed that increasing the anisotropy tends to lower $\Delta S_m$ (and similarly $\Delta T_{ad}$) by shifting them towards higher temperatures. Moreover, it is important to consider that for temperatures below the blocking temperatures, the spin-lattice relaxation slows down dramatically. Therefore, molecular spins tend to loose thermal contact with the lattice, as revealed for instance by specific heat experiments,[11,12,14,15] resulting in lower magnetic entropies and, consequently, lower MCE parameters. If one wants to search for better performances at low temperatures, preferred materials would then be magnetic clusters with large $S$ and small anisotropy.

In comparison with conventional magnetic nanoparticles, molecular magnets have the advantage of being mono-disperse. It is easy to show, indeed, that the size-distribution smears out the MCE response, yielding therefore to a lower efficiency.[78] The high-spin magnets $Fe_8$ and $Mn_{12}$ were the first molecule-based materials to be studied for enhanced MCE.[20] Because of their high-spin ground state $S = 10$, significant magnetic entropy changes were observed to occur around the blocking temperatures ($\approx 5$ K). Based on magnetization data, Tejada and co-workers observed a dependence of the maximum magnetic entropy change on the sweep rate of the applied field.[20] For a magnetic system, the total entropy content is given by $\Delta S_m/R = \ln (2S+1)$. It is therefore easy to show that $Fe_8$ and $Mn_{12}$, with their spin ground state $S = 10$ well-defined at low temperatures, cannot have values of $\Delta S_m$ larger than 12.5 and 11 J/kg K, respectively. These values are however difficult to observe because of the large anisotropies which are present in both systems. Thus, for the reasons already discussed, $Fe_8$ and $Mn_{12}$ may not be considered ideal materials in terms of MCE. The search of nearly isotropic molecular clusters led to the heterometallic octanuclear rings of type $Cr_7Cd$, that can be seen as an ordered arrangement of well-separated paramagnetic spins, having fast relaxations in the whole (experimental) temperature range. By means of specific heat experiments, the cooling efficiency of $Cr_7Cd$ was reported[21] to be in the temperature range below 2 K. In terms of MCE parameters, the only limitation of this material is given by the molecular spin value itself ($S = 3/2$) allowing not more than $-\Delta S_m = 5.1$ J/kg K.

The above-reported values of the magnetic entropy change expressed in J/kg K units, are significantly large and competing with the performance of more conventional intermetallic compounds based on lanthanide elements.[74] Can magnetic molecules do even better?

The recently[79] synthesized molecular cluster of type $Fe_{14}$, possessing a spin value as large as $S = 25$ and a relatively small cluster magnetic anisotropy seemed prone to be an ideal choice. A full investigation,[22] both experimentally and theoretically, of the magnetocaloric properties of this molecule is summarized in Figure 14. The specific heat data reveals the existence of a phase transition below $T_N = 1.87$ K to a long-range magnetically ordered state. This implies a relatively small cluster magnetic anisotropy, otherwise superparamagnetic blocking above $T_N$ should be observed. However, even a small anisotropy may become important for a $S$ as large as that of $Fe_{14}$. This is reflected, for instance, in the relative height of the transition peak at $T_N$ that apparently is a bit too small for such a large spin value, suggesting that a large portion of the magnetic entropy is not available for the ordering mechanism. The data measured above 20 K show a large increase, that we associate with the lattice contribution. From the specific heat data, we evaluate both $\Delta S_m$ and $\Delta T_{ad}$. The results, depicted in the lower panel of Fig. 14, show that $\Delta S_m$ is exceptionally large reaching the respectable value of 5.0 $R$ or equivalently 17.6 J/kg K for $\mu_0 \Delta H = 7$ T, and surprisingly exceeding even the entropy expected for a $S = 25$ spin system, that is $R \ln(2S+1) = 3.9\ R$. This apparent contradiction was resolved by modelling the exchange interactions between the single-ion spins inside the $Fe_{14}$ cluster. It was then found a high degree of frustration which led to associate the excess of experimental magnetic entropy with the presence of low-lying net spin states close in energy to the $S = 25$ ground state.[22a] For the temperature range below 10 K, the reported $\Delta S_m$ and $\Delta T_{ad}$ values are better than that of intermetallic compounds. One of the best representatives, for instance, is the recently[80] studied ($Er_{1-x}Dy_x$)$Al_2$ alloy that, for $x > 0.5$ concentrations and $T < 10$ K, presents a $-\Delta S_m$ which is at least 30% smaller than that of $Fe_{14}$. The temperature range of efficiency of $Fe_{14}$ ($T < 10$ K) is easily accessible in any research laboratory with liquid $^4$He. The $Fe_{14}$ molecular cluster turns out to be therefore a fine cooling material to achieve very-low temperatures by adiabatic demagnetization.

Refrigerant materials that are suitable for operations at high temperatures attract the interest of a large number of researchers working in the field of magnetic refrigeration.[74] Room-temperature magneto-cryocoolers are expected to be commercially available in a near future. Molecule-based magnetic materials did not enter in this contest so far. To search for molecular clusters with larger and larger anisotropies will not finally pay back for the reasons discussed above so one needs to find other strategies. Generally speaking, one may try to involve more degrees of freedom of a molecular system thus increasing the molar entropy involved. These degrees of freedom need, however, to be tuneable by an external magnetic field. An alternative route to follow is to strengthen the intermolecular magnetic correlations which ultimately will give rise to long-range magnetic order (the magnetic entropy change is maximized when the material is near its magnetic ordering temperature). In this respect, extended molecule-based systems like the Prussian blue analogues are of particular interest since magnetic phase transitions up to room temperature have been reported for this class of compounds.[64] Specific heat and magnetization experiments have already been performed on selected Prussian blue analogues, revealing encouraging magnetic entropy changes for temperatures above 200 K.[23]

## 8. Open questions and perspectives

Molecule-based magnetic materials have given new impetus to the subject of specific heat. Fundamental phenomena have been discovered or at least corroborated by specific heat measurements. Indeed, the selected examples here reviewed show that this subject of investigation is very successful if applied to this class of materials, and complementary to magnetization and spectroscopy techniques.

A charm of molecule-based materials resides on the variety of hierarchies of the magnetic interactions and symmetries of the magnetic structure that is not achievable in intermetallic compounds and probably not even in other physical systems. This makes them very appealing for magnetothermal investigations, since magnetic correlations, either short- or long-range, are promptly detected by a specific heat experiment. Mechanisms of magnetic order in these materials have indeed attracted much interest so far, especially in systems with low magnetic dimensionality,[81] in which the ordering is driven by exchange as well as dipolar interactions. The case of pure dipolar magnetism, so seldom encountered in Nature (see Section 6), is one of the most challenging, i.e., for the determination of the critical exponents of the specific heat at the phase transition. A further interesting topic is the possible occurrence of a quantum phase transition at very low temperatures.[72] It is not clear whether an anomaly in the specific heat and the corresponding critical exponent can be measured in these cases but certainly magnetothermal investigations may provide fundamental information on the low temperature phase diagram of ordered molecular spin systems. The difficulties are not negligible and at the moment the main constrains seem to be related to the size of the available crystals on one side and to the very low temperature needed.

The dynamics of a spin system embedded in a environment made of phonons, nuclear and electronic spins, will probably continue to attract much attention and some of the issues related to this topic are addressable by magnetothermal investigations as depicted by the examples previously discussed. Yet, since a fundamental role is played by phonons and nuclear spins, one may wonder whether in certain conditions it can be possible and even more convenient to monitor directly the phonon or the nuclear contributions to the specific heat in order to find anomalies on those sides too. For instance, it would be of interest to investigate whether a non canonical distribution of phonons, like in a phonon bottleneck, may give rise to detectable anomalies in the thermal properties of the electron spins. More in general, the study of thermal properties has well evidenced that little is still known on the energy spectrum and on the properties of the molecular lattice, so any theoretical or experimental effort in this direction is certainly worth.

In spite of the fact that molecule-based materials have been so far scarcely investigated for magneto-cooling applications, results have already shown that they represent a valid alternative to conventional materials, at least at low temperatures. Investigations are presently determining the MCE of these materials by *direct* magnetothermal methods,[82] and exploring possibilities of applications and different routes to enhance the MCE even further.

Because of their appeal, we expect future developments in the topics sketched above to be rapidly forthcoming.

## Acknowledgements

A special thank goes to J. C. Lasjaunias, P. Gandit, J. Chaussy and I. Sheikin (CNRS, Grenoble) and M. Novak who much contributed to the development of this field. This work was carried out within the framework of the EC-Network of Excellence "MAGMANet" (No. 515767), WorkPackages 2, 9 and 10.

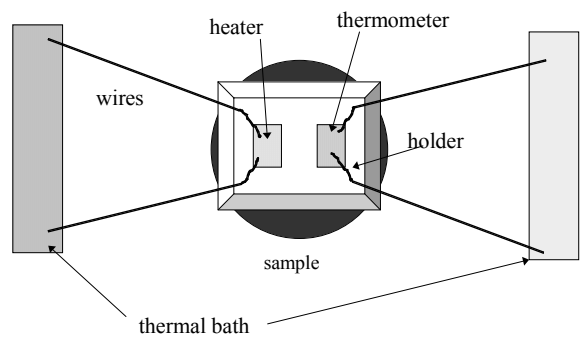

**Fig. 1** Scheme of an essential calorimeter.

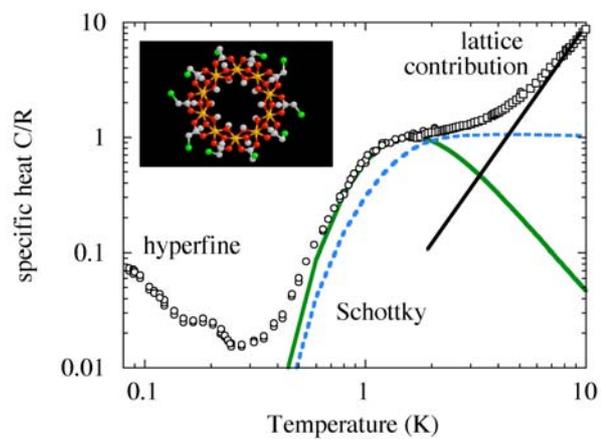

**Fig. 2** Low-temperature specific heat of $Fe_{10}$ ferric wheels. The lattice contribution dominates at high temperature and scales as $T^{2.69}$; the Schottky anomaly is clearly visible at low-$T$ and can be fitted to a simple model (green continuos curve) with two levels (a lower singlet and an upper triplet) separated by an energy gap of 4.56 K. If levels up to the fifth excited multiplet are considered and parameters are fixed by magnetic measurements, the blue dotted curve is obtained. At very low $T$, an upturn scaling as $T^{-2}$ is observed and ascribed to the hyperfine interaction with the neighbouring nuclei.

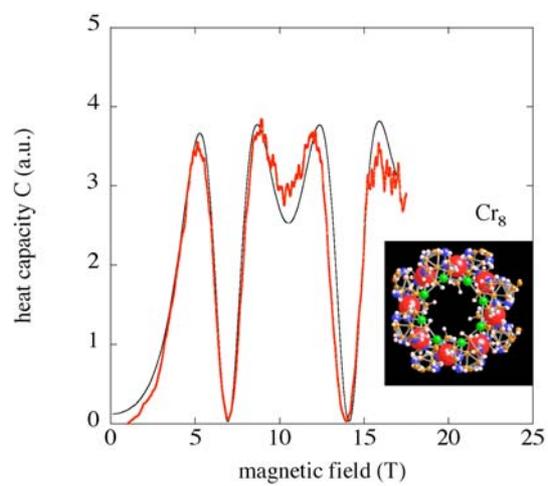

**Fig. 3** Magnetic field dependence of the isothermal heat capacity measured on a $Cr_8$ single crystals at 0.78 K. Maxima are observed when $\Delta = 2.5\ k_BT$, whilst the heat capacity vanishes at the crossing fields $B_{C1} = 6.9$ T and $B_{C2} = 14.0$ T.

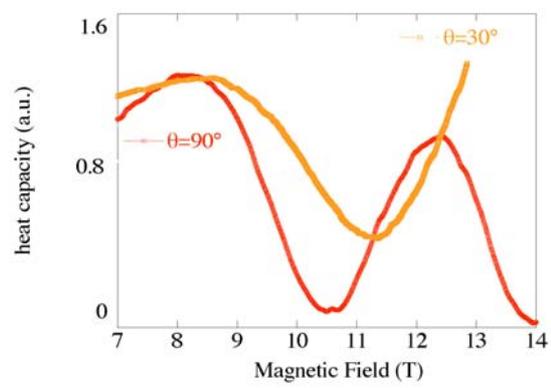

**Fig. 4** Heat capacity measured at the crossing field on a Cr$_7$Ni crystal (for θ = 90° and θ = 30°).

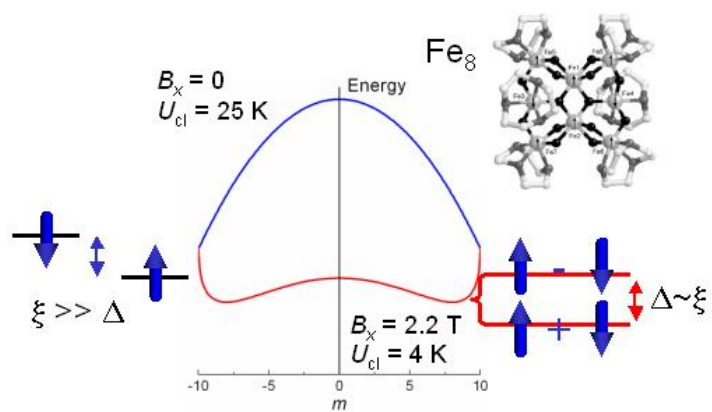

**Fig. 5** Classical potential energy $U_{cl} = -DS_z^2 - E(S_x^2 - S_y^2)$ of Fe$_8$ magnetic moments rotating in the $XZ$ plane. It was calculated in zero-field and in the presence of a magnetic field applied along the $X$ axis plane, with parameters $D = 0.292$ K, $E = 0.045$ K.[44] The quantum number $m$ gives the projection of the molecular spin along the anisotropy axis $Z$. Inset: Sketch of the Fe$_8$ magnetic core.

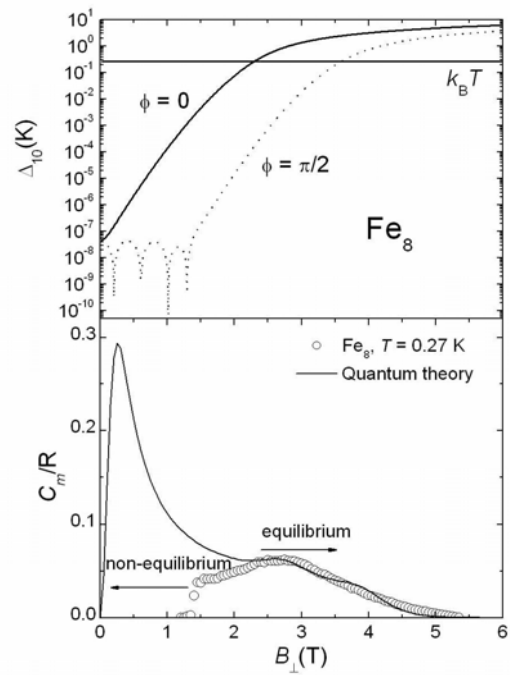

**Fig. 6** Top: Tunnel splitting of the magnetic ground state doublet of $Fe_8$ clusters, calculated for $D = 0.292$ K and $E = 0.045$ K as a function of the transverse magnetic field. Bottom: Specific heat of a sample of oriented $Fe_8$ microcrystals.[11] The line shows the equilibrium specific heat calculated by using the quantum energy levels of $Fe_8$ and a uniform distribution of easy axes orientations between 81° and 90°. The low-field contribution corresponds to those crystals that are not perfectly perpendicular to the magnetic field. It is not observed experimentally because $Fe_8$ spins are in thermal equilibrium only for $B_\perp > 1.7$ T.

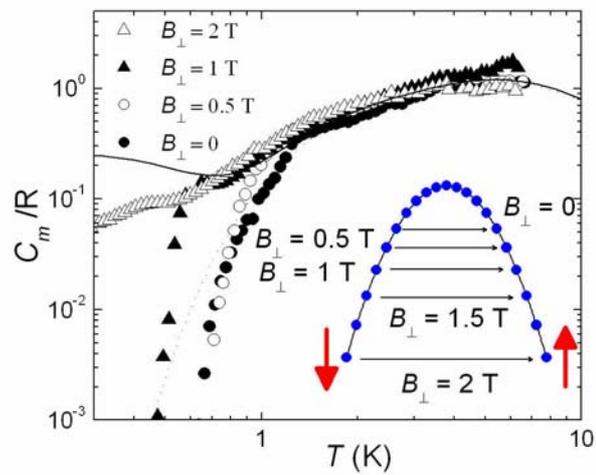

**Fig. 7** Temperature dependence of the magnetic specific heat of Fe$_8$ measured under different transverse magnetic fields. The solid line is the calculated $C_m^{eq}$.[11] The inset shows the states via which tunneling takes place predominantly as the field increases.

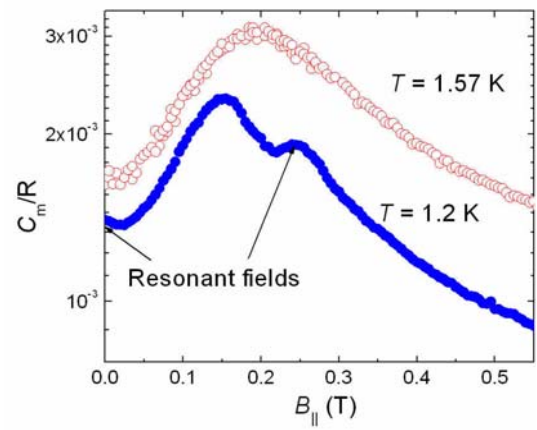

**Fig. 8** Magnetic specific heat of a sample of oriented powder of $Fe_8$ as a function of magnetic field applied parallel to the orientation's direction. The two temperatures shown are larger and smaller than the blocking temperature ≈ 1.3 K, respectively.

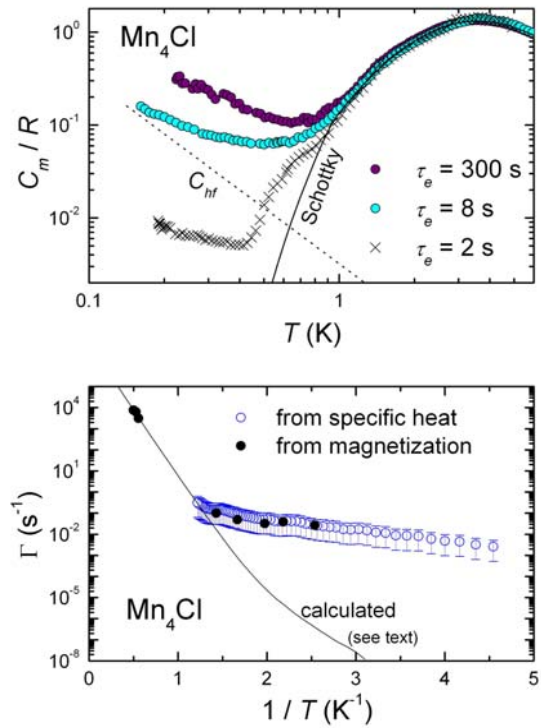

**Fig. 9** Top: Temperature-dependent magnetic specific heat of Mn$_4$Cl measured for $\tau$ = 2, 8, and 300 s (at $T$ = 0.4 K). Solid line is the Schottky contribution; dotted line is the expected nuclear contribution $C_{hf}$. Bottom: Spin-lattice relaxation rate of Mn$_4$Cl obtained: (●) from magnetic relaxation data;[59] (○) by fitting the $C_m$ vs $t$ data of the top panel (see Ref. [15]). Solid line is calculated for magnetic fields $B_x$ = 150 G and $B_z$ = 350 G.[54]

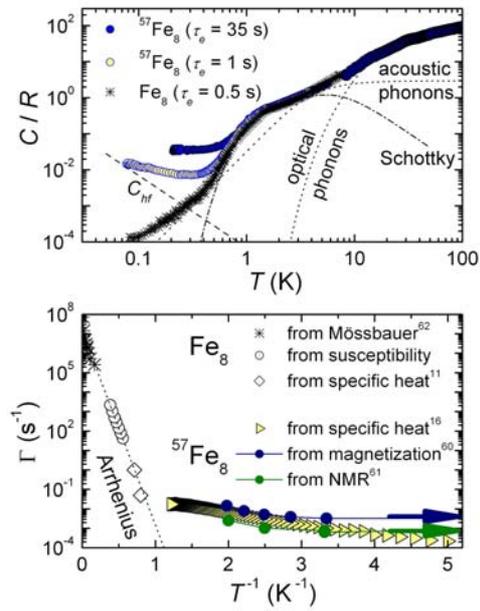

**Fig. 10** Top: Zero-field specific heat of non-oriented samples of standard Fe$_8$ and of $^{57}$Fe$_8$ as a function of temperature. For $^{57}$Fe$_8$, data for $\tau = 35$ s and 1 s (at $T = 0.2$K) are given, whereas for Fe$_8$, $\tau = 0.5$ s (at $T = 0.2$ K), as labeled. Drawn curves represent the different contributions to the specific heat.[16] Bottom: Spin-lattice relaxation rates of standard Fe$_8$ and $^{57}$Fe$_8$ obtained by different experimental techniques, as labeled. Dashed curve is the Arrhenius fit to the high-$T$ data. Arrows indicate low-$T$ limits deduced from magnetization[60] and NMR[61] data.

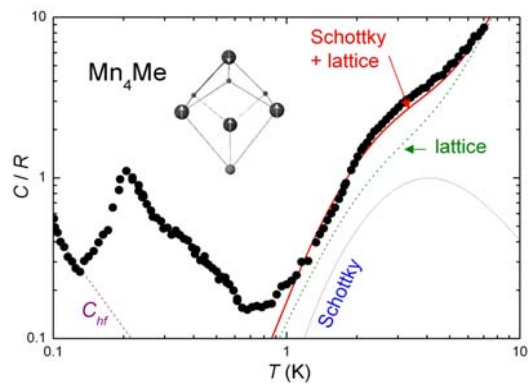

**Fig. 11** Temperature-dependence of the zero-field specific heat of Mn$_4$Me showing the transition to a long-range ordered state below $T_C$ = 0.21 K. Dashed line denotes the calculated hyperfine contribution $C_{hf}$; solid line, sum of lattice (dotted line) plus Schottky contributions. Inset: Sketch of the magnetic core.

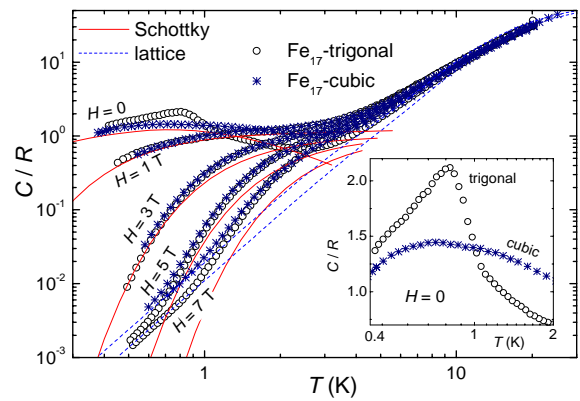

**Fig. 12** Temperature-dependent specific heat of $Fe_{17}$-trigonal and $Fe_{17}$-cubic, as labelled, for several applied-fields. Drawn curves are explained in the text. Inset: Magnification of the zero-field and low-temperature range showing the different ordering behaviours.

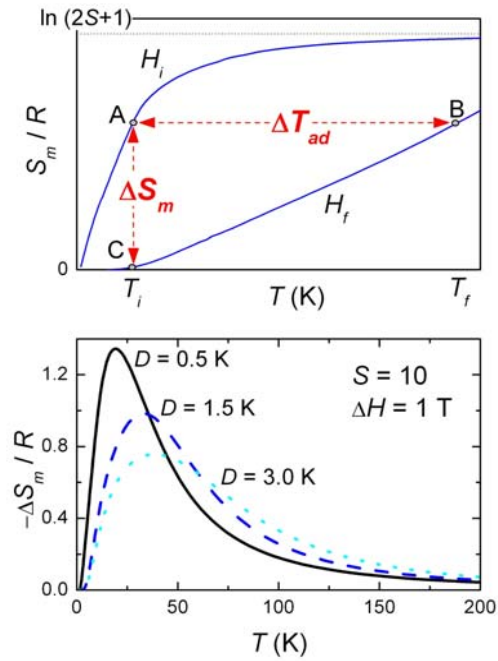

**Fig. 13** Top: The material, assumed to be a (super)paramagnet with spin $S$ per formula unit, is initially in state $A(T_i,H_i)$, at temperature $T_i$ and field $H_i$. Under adiabatic conditions (i.e. when the total entropy of the system remains constant during the magnetic field change), the magnetic entropy change must be compensated for by an equal but opposite change of the entropy associated with the lattice, resulting in a change in temperature of the material. That is, the *adiabatic* field change from $H_i$ to $H_f$ brings the system to state $B(T_f,H_f)$ with the temperature change $\Delta T_{ad} = T_f - T_i$ (horizontal arrow). If alternatively the magnetic field is *isothermally* changed to $H_f$ in a reversible process, the system goes to state $C(T_f,H_f)$ with the magnetic entropy change $\Delta S_m$ (vertical arrow). It is easy to see that if the magnetic change $\Delta H$ reduces the entropy ($\Delta S_m < 0$), then $\Delta T_{ad}$ is positive, whereas if $\Delta H$ is such that $\Delta S_m > 0$, then $\Delta T_{ad} < 0$. Bottom: Calculated $\Delta S_m$ for an isolated magnetic particle with $S = 10$ and varying axial anisotropy $D = 0.5$, 1.5 and 3.0 K upon a field change $\Delta H = (1 - 0)$ T.

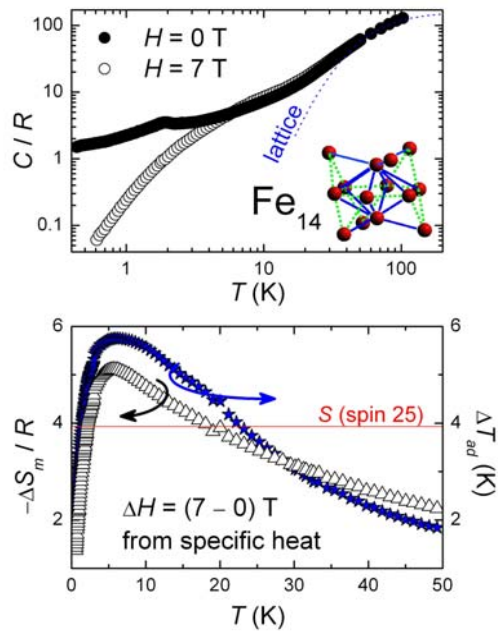

**Fig. 14** Top: Experimental specific heat of $Fe_{14}$ for applied magnetic field $H$ = 0 and 7 T. The dotted line is the high-$T$ lattice contribution. In the inset, sketch of the symmetric core containing fourteen exchange-coupled $Fe^{3+}$ ions. Bottom: magnetic entropy change $\Delta S_m$ and adiabatic temperature change $\Delta T_{ad}$ as obtained from the experimental data of the top panel. The solid line is the expected entropy for a spin $S$ = 25.